\documentclass[iop,apj,dvipdfmx]{emulateapj}

\shorttitle{Particle Acceleration in Quasi-parallel Shocks}
\shortauthors{Tsunehiko N. Kato}

\newcommand{\bvec}[1]{\mbox{\boldmath $#1$}}
\newcommand{\mel}{m_\mathrm{e}}
\newcommand{\mpr}{m_\mathrm{p}}
\newcommand{\omegape}{\omega_\mathrm{pe}}
\newcommand{\omegace}{\omega_\mathrm{ce}}
\newcommand{\lambdae}{\lambda_\mathrm{e}}
\newcommand{\lambdap}{\lambda_\mathrm{p}}
\newcommand{\omegacp}{\omega_\mathrm{cp}}
\newcommand{\vA}{v_\mathrm{A}}
\newcommand{\Ekin}{E_\mathrm{kin}}
\newcommand{\Te}{T_\mathrm{e}}
\newcommand{\Tp}{T_\mathrm{p}}
\newcommand{\Teq}{T_\mathrm{eq}}
\newcommand{\nez}{n_\mathrm{e0}}
\newcommand{\Ne}{N_\mathrm{e}}
\newcommand{\NPPC}{N_\mathrm{PPC}}

\newcommand{\colver}{}

\begin{document}

\title{Particle acceleration and wave excitation in quasi-parallel high-Mach-number collisionless shocks:
Particle-in-cell simulation}

\author{Tsunehiko N. Kato\altaffilmark{1}}
\affil{Department of Physical Science, Graduate School of Science, Hiroshima University}
\affil{1-3-1, Kagamiyama, Higashi-Hiroshima, Hiroshima 739-8526, Japan}
\altaffiltext{1}{Current address: Center for Computational Astrophysics,
National Astronomical Observatory of Japan,
2-21-1 Osawa, Mitaka, Tokyo 181-8588, Japan
}
\email{tsunehiko.kato@nao.ac.jp}

\begin{abstract}
We herein investigate shock formation and particle acceleration processes
for both protons and electrons
in a quasi-parallel high-Mach-number collisionless shock
through a long-term, large-scale particle-in-cell simulation.
We show that
both protons and electrons are accelerated in the shock
and that these accelerated particles generate large-amplitude Alfv\'{e}nic waves
in the upstream region of the shock.
After the upstream waves have grown sufficiently,
the local structure of the collisionless shock
becomes substantially similar to that of a quasi-perpendicular shock
due to the large transverse magnetic field of the waves.
A fraction of protons are accelerated in the shock
with a power-law-like energy distribution.
The rate of proton injection to the acceleration process
is approximately constant, and in the injection process,
the phase-trapping mechanism for the protons by the upstream waves
can play an important role.
The dominant acceleration process is a Fermi-like process
through repeated shock crossings of the protons.
This process is a `fast' process in the sense that
the time required for most of the accelerated protons to complete
one cycle of the acceleration process 
is much shorter than the diffusion time.
A fraction of the electrons is also accelerated by the same mechanism, and have
a power-law-like energy distribution.
However,
the injection does not enter a steady state during the simulation,
which may be related to the intermittent activity of the upstream waves. 
Upstream of the shock,
a fraction of the electrons is pre-accelerated before reaching the shock,
which may contribute to steady electron injection at a later time.
\end{abstract}


\section{Introduction}
Collisionless shocks, which are driven by
various violent phenomena throughout the universe,
are believed to be sites of particle acceleration.
In particular,
cosmic rays with energies below the knee energy (approximately $10^{15.5}$ eV)
are considered to be accelerated by shocks in supernova remnants (SNRs)
in our galaxy.
A number of X-ray observations have revealed 
synchrotron X-rays radiated from high-energy electrons around the shocks
in several young SNRs,
and these synchrotron X-rays are regarded as
evidence of electron acceleration around the shocks
to energies of approximately $10^{14}$ eV
\citep{Koyama95, Long03, Bamba03}.
Recent observations
have also revealed gamma rays associated with
the decay of neutral pions ($\pi^0$ mesons),
which occurs as a result of proton-proton collisions \citep{Ackermann13}.
These observations are regarded as direct evidence of
high-energy protons in the vicinity of the shocks.

One of the most plausible candidates for the acceleration process that acts in shocks
is first-order Fermi acceleration \citep{Drury83, Blandford87}.
In particular, first-order Fermi acceleration can explain the power-law energy spectrum for the accelerated particles
and the power-law index expected from the observations of cosmic rays.
However, the acceleration process that is actually operating in the SNR shocks
has not yet been determined, especially with respect to electron acceleration,
and further observational and theoretical investigations are needed.

Generally, it is considered that the acceleration efficiency
is strongly dependent on the shock velocity or the shock Mach number,
and the process would be more efficient for larger velocity
or higher Mach number shocks. For example, young SNRs have large shock velocities (approximately 1,000 to 10,000 km s$^{-1}$), and the corresponding Mach number is very large (approximately 100).
Young SNRs are believed to be efficient accelerators of particles.
In addition, the orientation of the background magnetic field upstream of the shock
can also have a significant influence on the efficiency of the particle acceleration process.
For example, the bipolar morphology of the emission region of non-thermal synchrotron X-rays
in the supernova remnant SN1006 is considered to be related to the orientation of the magnetic field in the interstellar medium around the remnant, which is upstream of the shock.
Recent observations of radio polarization by \citet{Reynoso13} suggest that the electron acceleration is efficient when the direction of the ambient magnetic field is approximately parallel to the shock normal. Hence, the particle acceleration process can be more efficient in quasi-parallel shocks than in quasi-perpendicular shocks. Here, shocks in which the upstream magnetic field lies along the shock normal are referred to as parallel shocks, and shocks in which the upstream field lies perpendicular to the shock normal are referred to as perpendicular shocks.

In the heliosphere,
collisionless shocks (e.g., Earth's bow shock, interplanetary shocks associated with coronal mass ejections,
solar wind termination shocks, etc.) are also formed.
These shocks have been investigated through
a number of direct \textit{in situ} observations by spacecraft:
For example, the Earth's bow shock \citep{Burgess12}
and those associated with solar energetic particles and energetic storm particle events \citep{Lee12}.
In particular, these observations indeed showed
that quasi-parallel shocks can be efficient ion accelerators
and their efficiency is generally dependent on the shock strength \citep{Reames00}.
The wave generation in the upstream region of shocks
by reflected energetic diffuse ions are also observed in Earth's bow shock
\citep{Hoppe81, Burgess05}.
Although the Alfv\'{e}n Mach numbers of these shocks are generally smaller
than those of SNR shocks,
recently,
a quasi-parallel collisionless shock at very high Mach number of $\sim100$
have been observed in Saturn's bow shock by Cassini spacecraft \citep{Masters13}.
This observation also showed an efficient electron acceleration there,
suggesting that quasi-parallel shocks at very high Mach numbers
can be efficient electron accelerators, too.

The acceleration processes in shocks have also been investigated through numerical simulations.
Recent large-scale hybrid simulations, in which protons are treated as discrete particles whereas electrons are approximated as a massless fluid, have shown that protons are accelerated efficiently in quasi-parallel shocks with a power-law-like energy distribution \citep{Giacalone04,Sugiyama11,Gargate12,Caprioli14}.
The acceleration process observed in these studies 
is often a `fast' process in the sense that the accelerated protons cross the shock front
back and forth repeatedly within a 
much shorter timescale than that of the diffusive motion.
Another prominent feature of these particle accelerating shocks
is the wave excitation in the upstream region of the shock
by the accelerated protons,
which would be a similar processes to those observed in quasi-parallel Earth's bow shocks
by \textit{in situ} observations.
The amplitude of these waves can be even larger than the strength
of the background magnetic field. These large-amplitude upstream waves can strongly influence both the particle acceleration process and the shock structure.

Hybrid simulations, however, cannot deal with the kinetic dynamics of electrons, especially non-thermal electron acceleration. In the present paper, we investigate the particle acceleration process for both protons and electrons in high-Mach-number quasi-parallel shocks, as well as the shock formation process and structure, taking into account the electron dynamics self-consistently through particle-in-cell (PIC) simulations in which both protons and electrons are treated as discrete particles. For protons to be accelerated to sufficiently high energies,
long-term (typically hundreds of proton gyro-time) and therefore large-scale simulations are required. In order for the calculation time to be as long as possible,
we carry out a large-scale one-dimensional simulation.
In Section 2, we describe the simulation model.
The primary results of the simulation are shown in Section 3.
The obtained results are discussed in Section 4, and conclusions are presented  in Section 5.

\section{Method}
In order to investigate the particle acceleration process
in quasi-parallel collisionless shocks in an electron-proton plasma,
we carry out a large-scale numerical simulation.
The simulation code is a one-dimensional electromagnetic
particle-in-cell code with one spatial dimension and
three velocity dimensions (1D3V)
that was developed based on a standard method described by
\citet{Birdsall}.
The basic equations are the Maxwell's equations and
the (relativistic) equation of motion of particles.
In the following,
we take the $x$-axis as the one-dimensional direction.

In the simulation,
a collisionless shock is driven according to the ``injection method''.
There are two conducting rigid walls at both ends of the simulation box.
These walls reflect both incident particles and electromagnetic waves specularly.
Initially, the plasma is moving in the $+x$-direction at a bulk velocity $V$.
Both electrons and protons are loaded uniformly
in the region between the two walls
with an average velocity of $V$ plus the thermal velocity,
where
the temperatures of the electrons and protons are initially set to be equal.
As the plasma is magnetized,
the plasma convects the ordered background magnetic field $\bvec{B}_0$.
Since the electric field should vanish in the plasma rest frame,
the motional electric field $\bvec{E}_0 = -\bvec{V} \times \bvec{B}_0$
appears in the simulation frame, in which the plasma is moving.
In the early stage of the simulation,
the particles that collide with the wall on the $+x$ side are reflected specularly
and then interact with the incoming plasma.
This interaction causes some instabilities
and eventually leads to the formation of a collisionless shock.
The frame of the simulation is the rest frame of the shock downstream plasma
(hereinafter, the downstream rest frame).
Thus,
in the simulation,
we observe the propagation of the collisionless shock
in the $-x$-direction in the downstream rest frame.

In the following,
we take $\omegape^{-1}$ as the unit of time
and $\lambdae = c\omegape^{-1}$ as the unit of length,
where $\omegape = (4\pi \nez e^2/\mel)^{1/2}$
is the electron plasma frequency
defined for the far upstream plasma number density $\nez$
with electron mass $\mel$ and magnitude of the electron charge $e=|e|$.
The units of the electric and magnetic fields
are given by $E_* = B_* = c(4\pi n_\mathrm{e0} \mel)^{1/2}$.

\section{Results}
\label{sec:results}
We carry out a large-scale one-dimensional PIC simulation under the following conditions.
We use a reduced proton mass of $\mpr = 30 \mel$
for which the proton inertial length is given by $\lambdap = 5.48\lambdae$.
The number of spatial grids is $N_x = 2.5\times10^6$,
and there are $\NPPC=160$ super-particles per cell per species.
The physical dimension of the simulation box is
$L_x = 1.5\times 10^5 \lambdae = 2.74\times 10^4 \lambdap$.
The size of a cell is thus $\Delta x = 0.06\lambdae$. 
The time step is taken to be $\Delta t = 0.05 \omegape^{-1}$.

The initial bulk velocity of the plasma is given by $V=0.37c$ in the $+x$-direction.
The corresponding Lorentz factor is given by $\Gamma = 1.08$,
and thus the relativistic effect is not significant for the shock formation.
The ordered background magnetic field is set on the $x$-$y$ plane, and,
in order to study a quasi-parallel shock,
the angle between it and the $x$-axis (i.e., the shock normal)
is taken to be $\Theta_0 = 30^\circ$ in the downstream rest frame.
Hence, the motional electric field lies in the $-z$-direction.
The strength of the background magnetic field $B_0 = |\bvec{B}_0|$ is set so that
$B_0/B_* = \omegace / \omegape = 8.847\times 10^{-2}$,
where $\omegace = eB_0/\mel c$ is the electron cyclotron frequency.
The proton gyro frequency is given by $\omega_\mathrm{cp} = 2.95\times10^{-3} \omegape$.
The Alfv\'{e}n velocity in the upstream region is thus given by $\vA = 1.59\times10^{-2}c$.
In terms of the magnetization,
$\sigma \equiv B_0^2/4\pi \nez (\mel + \mpr) V^2$,
which corresponds to $\sigma = 1.9 \times 10^{-3}$,
i.e., it is very weakly magnetized.
The plasma beta parameters defined for electrons and protons are both set to be $0.5$,
so that the total plasma beta becomes unity
($\beta = \beta_\mathrm{e} + \beta_\mathrm{p} = 1$)
and the initial temperatures of electrons and protons are determined accordingly
($\Te = \Tp = 1.96 \times 10^{-3} \mel c^2$).

\subsection{Structure of Collisionless Shock}
\label{subsection:shock_structure}
\begin{figure}
\plotone{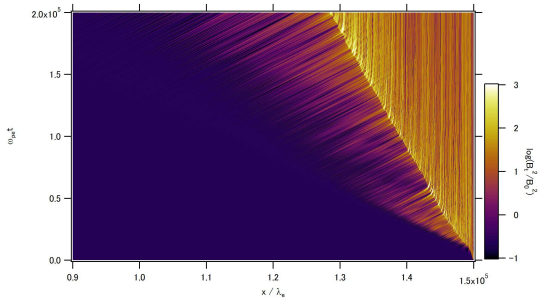}
\caption{
Time evolution of the energy density of
the transverse magnetic field $B_t = (B_y^2 + B_z^2)^{1/2}$.
The horizontal and vertical axes indicate the $x$-coordinate and the time, respectively.
The value of $\log_{10}(B_t^2/B_0^2)$ is shown in grayscale (color).
A collisionless shock is formed on the right-hand side and propagates to the left.
Large-amplitude ($B_t \gg B_0$) waves are evident
upstream (left-hand side) of the shock.
\colver
}
\label{fig:evol_logUB}
\end{figure}

Figure \ref{fig:evol_logUB} presents the evolution of the magnetic energy density
of the transverse components, $B_t = (B_y^2 + B_z^2)^{1/2}$, normalized by that of
the background field $B_0^2/8\pi$.
A collisionless shock is observed to form
in the vicinity of the wall located at $x=1.5\times 10^5 \lambdae$
and then propagate to the left at an approximately constant speed.
The shock speed is obtained as approximately $-0.1c$
in the downstream rest frame (namely, the simulation frame)
and approximately $-0.46c$ in the upstream rest frame.
The corresponding Alfv\'{e}n Mach-number is given by $M_A \sim 28$.%
(Note that although this Mach number would be generally regarded as `high' Mach number,
it is still small compared with the typical values for SNR shocks $M_A \sim 100$.)
It is also evident in this figure that there exist large-amplitude ($B_t \gg B_0$) waves in the upstream region (left-hand side) of the shock.
These waves appear around $\omegape t \sim 1.5 \times 10^4$, and
the region in which they exist, which is sometimes referred to as the foreshock region,
extends upstream with time.
This excitation of the waves in the upstream region
can be attributed to the appearance of the energetic particles \citep{Bell78}.
Indeed, excitation is commonly observed in numerical simulations
in which a fraction of the particles is accelerated in the shock efficiently
under the quasi-parallel condition,
for example,
in non-relativistic hybrid simulations \citep{Giacalone04,Caprioli14}
and relativistic PIC simulations \citep{Sironi11}.

\begin{figure}
\plotone{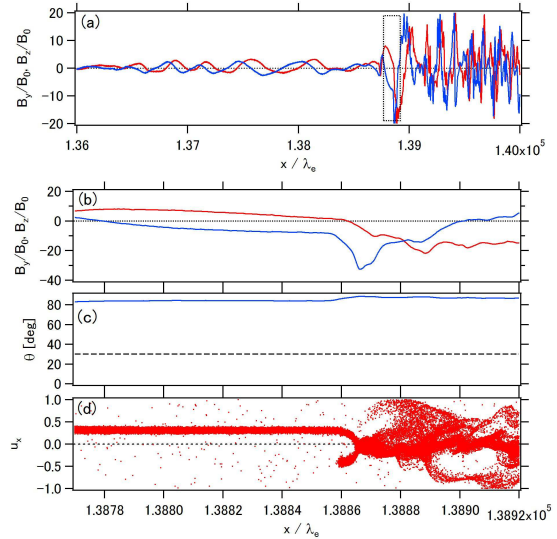}
\caption{
(a) Magnetic fields $B_y$ (black (red)) and $B_z$ (gray (blue)) around the shock
at $\omegape t = 1\times 10^5$ normalized by the background field $B_0$. 
(b) Close-up of the region enclosed by the dotted box in (a).
(c) Angle of the local magnetic field to the shock normal ($x$-axis).
(d) $x-u_x$ phase-space plot of the protons,
where $u_x$ denotes the $x$-component of the particle 4-velocity.
\colver
}
\label{fig:perp_shock}
\end{figure}

Figure \ref{fig:perp_shock}(a) shows
the profiles of the magnetic field components $B_y$ and $B_z$
around the shock at $\omegape t = 1\times 10^5$,
where the shock front is located at $x \sim 1.3886\times10^5 \lambdae$.
Large-amplitude waves occur upstream of the shock,
and the magnetic fields of the waves
are much larger than the background field $B_0$
in the vicinity of the shock.
Figure \ref{fig:perp_shock}(b) shows a close-up of the region
indicated by the dotted box in Fig.~\ref{fig:perp_shock}(a).
The transverse magnetic fields are dominant around the shock.
As shown in Fig.~\ref{fig:perp_shock}(c),
the local magnetic field and the shock normal (i.e., the $x$-axis)
are approximately perpendicular.
Since the shock structure is much smaller than the typical wavelength
of the upstream waves, the shock experiences an almost uniform perpendicular magnetic field.
Therefore, the local shock structure itself becomes essentially that of the \textit{quasi-perpendicular} shock.
Indeed, Fig.~\ref{fig:perp_shock}(d) shows that
some of the incoming upstream protons are reflected at the shock front,
which is a well-known characteristic of quasi-perpendicular shocks.
Due to this strong perpendicular magnetic field,
the local Alfv\'{e}n Mach number becomes $M_A \sim 3.6$.
Thus, the shock itself is locally not a high-Mach-number shock.

\subsection{Wave Excitation in the Upstream Region}
\begin{figure}
\plotone{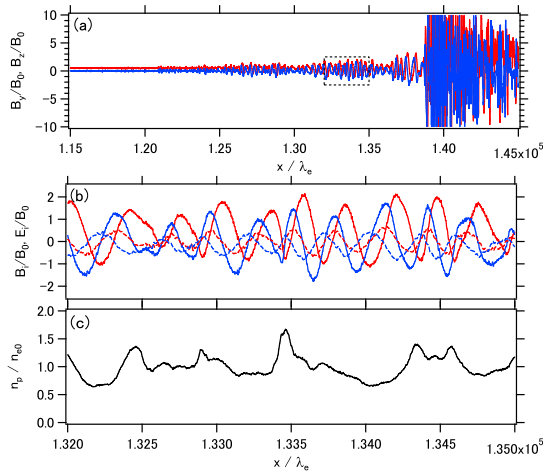}
\caption{
(a) Magnetic fields around the shock
at $\omegape t = 1\times 10^5$, as in Fig.~\ref{fig:perp_shock}(a). 
(b) Close-up of the region enclosed by the dotted box in (a).
The normalized electric fields $E_y/B_0$ and $E_z/B_0$
are also shown by the dashed curves.
(c) Number density of protons $n_\mathrm{p}$
normalized by the number density far upstream $\nez$.
\colver
}
\label{fig:profile_EB}
\end{figure}

Figure \ref{fig:profile_EB}(a) shows
the profiles of the magnetic field
around the shock, as in Fig.~\ref{fig:perp_shock}(a).
Figure \ref{fig:profile_EB} (b) shows a close-up of the region
$1.32\times10^5 < x/\lambdae < 1.35\times10^5$
indicated by the dotted box in Fig.~\ref{fig:profile_EB}(a) together with the electric fields.
The generated waves are monochromatic rather than turbulent,
as is clear from the figure.
The wavelengths of these waves
are typically $\lambda \sim 200\lambdae - 400\lambdae$,
which is comparable to or somewhat larger than the gyro-radius of the protons
reflected at the shock
(defined for the background field strength $B_0$).
The wavelength satisfies the condition
$\lambda \gg 2\pi \vA/\omegacp \sim 34 \lambdae$,
where $\omegacp = eB_0/\mpr c$ is the cyclotron frequency of the upstream protons,
and so are regarded as Alfv\'{e}nic electromagnetic waves.
Although the waves actually propagate obliquely with respect to the background field,
the structures of the electric and magnetic fields of the waves
are essentially the same as the structure of right-hand circularly polarized waves
and are naturally explained to be generated
by the resonant mode instability \citep{Winske84}.
The number density of the incoming upstream plasma
is also modulated because of the existence of the waves, as shown in Fig.~\ref{fig:profile_EB}(c).
This compressive behavior can be explained as a non-linear effect or
a feature of obliquely propagating unstable modes \citep{Gary81}.

\begin{figure}
\plotone{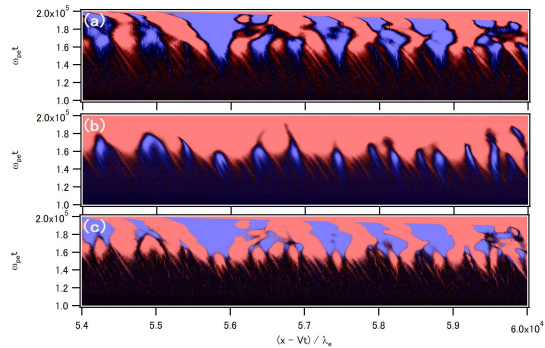}
\caption{
Evolution of upstream waves observed in the upstream rest frame.
(a) Fluctuation in the proton number density $\delta n_\mathrm{p}$.
(b) Fluctuation in the magnetic pressure $\delta P_B$.
(c) The product $\delta n_\mathrm{p} \times \delta P_B$, which shows the correlation of
the density fluctuation and the magnetic fluctuation.
In all panels, positive values are shown in white (red),
and negative values are shown in black (blue).
The horizontal coordinate is $x - Vt$,
where $V$ is the upstream flow velocity
measured in the simulation frame.
\colver
}
\label{fig:upstream_wave_evol}
\end{figure}

Figure \ref{fig:upstream_wave_evol} shows the features of the upstream waves
observed in the upstream rest frame.
The fluctuation in the proton number density $\delta n$,
the fluctuation in the magnetic pressure $\delta P_B$, 
and the product of these fluctuations
are presented in Figs.~\ref{fig:upstream_wave_evol}(a), \ref{fig:upstream_wave_evol}(b), and \ref{fig:upstream_wave_evol}(c), respectively.
In Fig.~\ref{fig:upstream_wave_evol}(a),
the shock front is visible as the slightly inclined horizontal discontinuity
at nearly $\omegape t \sim 2\times10^5$
and
the region below this shock front is the upstream region of the shock.
Note that when the amplitude of the waves is small
($\omegape t \lesssim 1.5 \times 10^5$ in this figure),
the waves propagate in the upstream ($-x$) direction,
which is consistent with the explanation, which indicates that the waves are generated by the resonant mode instability.
In addition, there is a positive correlation between
the density fluctuation and the magnetic fluctuation.
This can also be explained as a feature of the oblique mode
\citep{Gary81}.
In the nonlinear regime,
the waves are almost at rest in the upstream rest frame and grow in wavelength and amplitude.

\subsection{Particle Acceleration}
\label{subsec:particle_acceleration}
\begin{figure}
\plotone{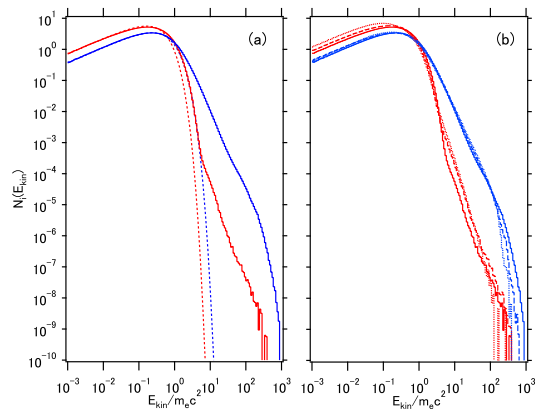}
\caption{
(a) Energy spectra of protons (black (blue) solid curve) and electrons (gray (red) solid curve) 
in the downstream region of the shock
obtained from the simulation at $\omegape t = 2 \times 10^5$.
Thermal Maxwellian distributions for protons with temperature $\Tp=0.46 \mel c^2$ and
for electrons with temperature $\Te=0.24 \mel c^2$
are also shown by 
the black (blue) dashed curve and the gray (red) dashed curve, respectively.
(b) Development over time of the energy spectra.
The spectra for $\omegape t = 5\times10^4$ (dotted curves),
$1.0 \times 10^5$ (dashed curves), and
$2.0 \times 10^5$ (solid curves) are shown.
\colver
}
\label{fig:hist_d}
\end{figure}

Figure \ref{fig:hist_d}(a) presents the energy spectra
of the protons and the electrons in the downstream region of the shock
at $\omegape t = 2 \times 10^5$
together with the fitted thermal Maxwellian distributions.
Here,
$\Ekin = (\gamma-1)mc^2$ denotes
the kinetic energy of particles of mass $m$ and Lorentz factor $\gamma$
measured in the downstream rest frame.
Note that the bulk kinetic energy of the protons
and that of the electrons of the incoming plasma in the upstream region
are given as approximately $2.3 \mel c^2$ and approximately $7.6 \times 10^{-2} \mel c^2$, respectively.
In this figure,
high-energy and non-thermal populations with power-law-like distributions exist
not only in protons (for $\Ekin \gtrsim 3 \mel c^2$) but also in electrons (for $\Ekin \gtrsim 5 \mel c^2$).
The power-law indices obtained from these portions of the distributions
are approximately $2.4$ for protons and approximately $3.0$ for electrons. 
However,
the acceleration efficiency of the electrons
is clearly low compared with that of the protons.
The amount of non-thermal electrons is approximately 
two orders of magnitude smaller than that of the non-thermal protons. 
Moreover,
the highest energy of the electrons ($\Ekin \sim 400 \mel c^2$)
is smaller than that of the protons ($\Ekin \sim 1,000 \mel c^2$).

The number densities and temperatures of the Maxwellian distributions
for fitting the thermal components in Fig.~\ref{fig:hist_d}(a)
are given by
$n_\mathrm{p,th}=3.3 \nez$ and $\Tp=0.46 \mel c^2$ for protons
and $n_\mathrm{e,th}=3.5 \nez$ and $\Te=0.24 \mel c^2$ for electrons,
while the total number densities (including the non-thermal components)
are given by $n_\mathrm{p} \sim n_\mathrm{e} \sim 4.4 \nez$.
Here, since a fraction of the thermal electrons can be relativistic,
we adopted a relativistic Maxwellian (or J\"{u}ttner-Synge) distribution \citep[c.f.][]{Landau}.
The temperature ratio $\Tp/\Te \sim 1.8$ is much smaller than the mass ratio $30$
indicating efficient (but not complete) energy exchange between the protons and electrons in the shock.
On the other hand,
the equilibrium temperature obtained
if the upstream bulk kinetic energy of the plasma
is fully and equally converted into the thermal kinetic energy of the protons and the electrons
downstream is given by $\Teq = 0.79 \mel c^2$.
Therefore, both temperatures are lower than the equilibrium temperature.
This is attributed to the fact that
a fraction of energy is transferred to the non-thermal particles
through the particle acceleration process.
Note that the fitting by the Maxwellian distribution
is not satisfactory for the electrons.
This can be improved
by adding a high-temperature thermal component with a temperature of approximately $0.46 \mel c^2$
to the first Maxwellian component.
As shown later,
a fraction of the incoming electrons are accelerated in the upstream region
before arriving at the shock front
and these `pre-accelerated' electrons
would result in a high-temperature component.

Figure \ref{fig:hist_d}(b) shows the development over time of the energy spectra
in the shock downstream region. 
For protons, the high-energy power-law-like portion extends over time,
which is an expected behavior for the Fermi-like acceleration process with
a constant injection of seed particles.
However, this is not the case for electrons.
The amount of non-thermal electrons decreases with time.
As will be shown later,
this is because the injection of the electrons to the acceleration process
is not constant but rather occurs only at a particular time interval.

\begin{figure}
\plotone{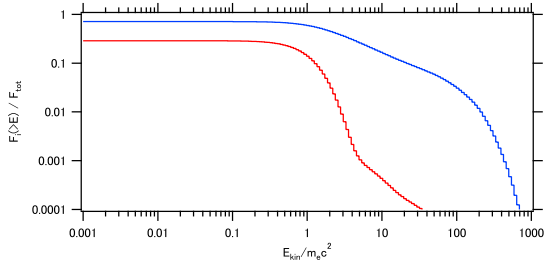}
\caption{
Cumulative energy spectra of the protons (black (blue)) and the electrons (gray (red))
in the downstream region of the shock at $\omegape t = 2 \times 10^5$
normalized by the mean total energy density.
\colver
}
\label{fig:cum_hist_d}
\end{figure}

Figure \ref{fig:cum_hist_d} shows  the cumulative energy distributions
defined by
\begin{equation}
	F_i(E) \equiv \int_E^\infty N_i(E') E' dE'
\end{equation}
in the downstream region at $\omegape t = 2 \times 10^5$ normalized by the mean total energy density,
where $i=\mathrm{e}, \mathrm{p}$, and
$N_i(E)$ is the energy spectrum for species $i$ shown in Fig.~\ref{fig:hist_d}.
The mean total energy density is given by $F_\mathrm{tot}=F_e(0)+F_p(0)$.
The total energy ratio of protons to electrons is approximately $2.5$.
The non-thermal protons and the non-thermal electrons contain approximately $10\%$ and approximately $0.1\%$, respectively of the total particle energy downstream.
These ratios are, of course, time dependent
and increase as the number of non-thermal particles
is increased through the acceleration process.

\subsubsection{Acceleration of Protons}
\label{subsubsec:proton_acceleration}
\begin{figure}
\plotone{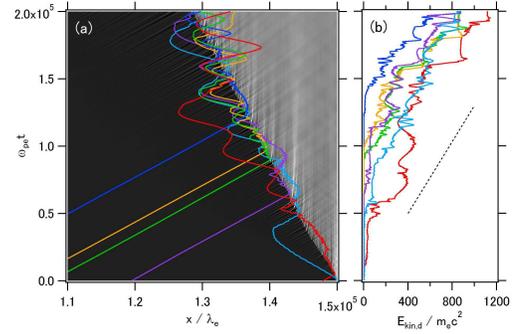}
\caption{
Acceleration histories of the highest-energy protons.
The $x-t$ trajectories are shown in (a).
The $\Ekin-t$ histories are shown in (b),
in which the kinetic energies $\Ekin$ are measured in the downstream rest frame.
The dotted line denotes the case of linear acceleration
at a rate of $d \Ekin/dt \sim 7.5\times10^{-3} (\mel c^2 \omegape)$.
\colver
}
\label{fig:accel_history_p}
\end{figure}

Figure \ref{fig:accel_history_p} presents the acceleration histories of the six protons
that are accelerated to the highest energies in the simulation.
The trajectories of these protons on the $x-t$ plane are shown in Fig.~\ref{fig:accel_history_p}(a)
and on the $\Ekin-t$ plane in Fig.~\ref{fig:accel_history_p}(b),
where $\Ekin$ denotes the kinetic energy measured in the downstream rest frame.
All of these protons are accelerated
upon repeatedly crossing back and forth across the shock
to energies up to $\Ekin \sim 1,000 \ \mel c^2$.
While the acceleration process is essentially a stochastic process,
these protons are, on average, accelerated linearly with time.
The average acceleration rate is roughly given by
$d \Ekin/dt \sim 7.5\times10^{-3} (\mel c^2 \omegape)$,
which is indicated by the dotted line in Fig.~\ref{fig:accel_history_p}(b).

\begin{figure}
\epsscale{0.8}
\plotone{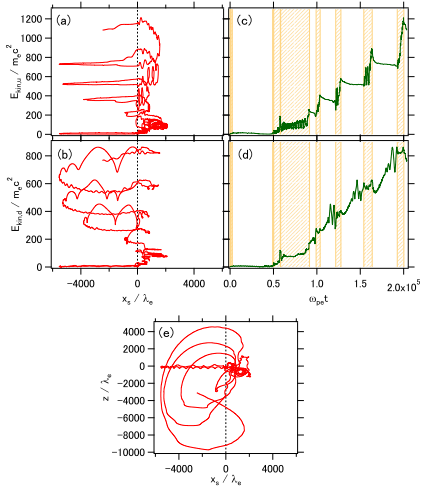}
\caption{
Representative acceleration history of protons.
(a) and (b) show the acceleration history on the $x_s-\Ekin$ plane,
where $x_s$ is the coordinate for which
the shock front is fixed at the origin.
The kinetic energy $\Ekin$
is measured in the upstream and
downstream rest frames.
(c) and (d) show those on the $t-\Ekin$ plane.
The shaded regions denote the time intervals
in which the proton remains in the downstream region and
the non-shaded regions denote the time intervals in which proton remains in the upstream region.
The trajectory on the $x_s-z$ plane is shown in (e).
\colver
}
\label{fig:particle_history_p}
\end{figure}

Figure \ref{fig:particle_history_p} shows a representative acceleration history of the protons.
In Figs.~\ref{fig:particle_history_p}(a) and \ref{fig:particle_history_p}(b),
the energy histories of the proton are presented on the $x_s-\Ekin$ plane,
where the kinetic energy $\Ekin$ is measured in the upstream rest frame in Fig.~\ref{fig:particle_history_p}(a)
and in the downstream rest frame in Fig.~\ref{fig:particle_history_p}(b), respectively.
Here, $x_s$ is the coordinate in which the shock front is fixed at the origin $x_s=0$
and given by $x_s(t) = x(t) - x_{sh}(t)$,
where $x(t)$ is the particle position and $x_{sh}(t)$ is the shock position,
both measured in the downstream rest frame.
As shown in Fig.~\ref{fig:accel_history_p},
the proton is accelerated by repeatedly crossing the shock.
Figures \ref{fig:particle_history_p}(c) and \ref{fig:particle_history_p}(d) show the variation of the corresponding particle kinetic energies over time.
Figures \ref{fig:particle_history_p}(a) through \ref{fig:particle_history_p}(d) show that the energy of the particle is almost conserved while it remains in one of the regions
if the energy is measured in the plasma rest frame in that region.
Therefore, the acceleration process is not a resonant process
but is essentially Fermi acceleration in the sense that particles are accelerated via repeated elastic scatterings off the scattering centers at different velocities.
The proton's trajectory on the $x_s-z$ plane is shown in Fig.~\ref{fig:particle_history_p}(e).
The trajectory is approximately a simple gyro-orbit
and is unlike that of the diffusive motion.
The non-diffusive features of the acceleration process are also observed
in recent simulations \citep{Sugiyama11, Sironi11}.
Furthermore, there is no apparent average drift in the $z$-direction,
namely the direction of the motional electric field,
suggesting that the acceleration process is not the shock drift acceleration.

\begin{figure}
\plotone{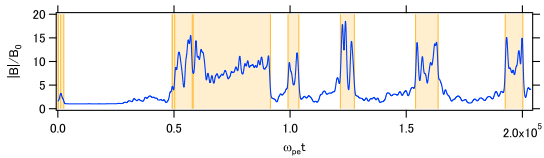}
\caption{
Local strength of magnetic field at the particle position normalized by the background field $B_0$
as a function of time for the same proton as in Fig.~\ref{fig:particle_history_p}.
The shaded regions indicate that the proton is in the downstream region.
\colver
}
\label{fig:Babs_history_p}
\end{figure}

Figure \ref{fig:Babs_history_p} shows the history of the local magnetic field strength
at the particle position
normalized by the upstream background field, $|\bvec{B}|/B_0$,
for the same proton, as shown in Fig.~\ref{fig:particle_history_p}.
After the large-amplitude waves are well developed in the upstream region
($\omegape t\gtrsim 3\times10^4$),
the strength of the local magnetic field is at least doubled in both the upstream and
downstream regions.
Accordingly, the typical gyration time $\tau_B= 2\pi \omega_{c}^{-1}$,
where $\omega_c = e |\bvec{B}| / \gamma \mpr c$ is the local proton cyclotron frequency,
becomes approximately half in each region,
shortening the acceleration timescale.

\begin{figure}
\plotone{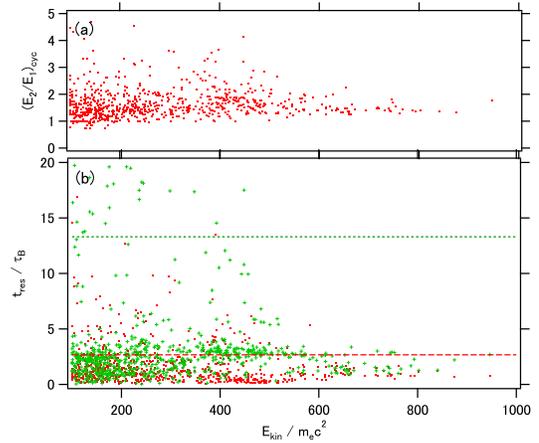}
\caption{
(a) Energy gain factor per cycle of shock crossing, $E_2/E_1$,
where $E_1$ and $E_2$ are the kinetic energies of the particle
before and after the cycle, respectively,
as a function of the kinetic energy before the crossing,
$\Ekin = E_1$,
for a sample of 1,000 protons that are accelerated to energies higher than $300 \mel c^2$
at the end of the simulation $\omegape t = 2\times 10^5$.
(b) Residence times in the upstream region (black (red) dots) and
the downstream region (gray (green) crosses) for one cycle.
The residence times in the upstream and
downstream regions derived from the Bohm diffusion model
are also shown by the dashed and dotted lines,
respectively. (See the text for details.)
\colver
}
\label{fig:crossing_p}
\end{figure}

Figure \ref{fig:crossing_p}(a) presents the energy gain factors
per cycle of the shock crossing of the protons
as a function of the energies before the shock cycle
for a sample of 1,000 protons
that are accelerated to energies higher than $300 \mel c^2$
at $\omegape t = 2\times 10^5$.
The factors are distributed around approximately 1 to 2, and
for high energies
they converge to approximately 1.4.
The particles for which the energies more than double during a half cycle
are accelerated within that region rather than upon shock crossing,
although these acceleration processes are not dominant.
There also exists a small fraction of particles that lose energy.
Figure \ref{fig:crossing_p}(b) shows the residence time of the particles in the upstream region and that in the downstream region for one cycle normalized by the average gyration time $\tau_B$.
Here, the average gyration time is taken
over each half cycle of the shock crossing.
Most of the protons return to the shock front
within times on the order of the average gyration time $\tau_B$.

For one cycle of the shock crossing of the particles, the residence time until return to the shock front in each region for diffusive motion is approximately given by \citep[cf.][]{Kato03}
\begin{equation}
	t_\mathrm{res} = \frac{4}{3} \frac{c}{|V|} \tau_0,
\end{equation}
where $\tau_0$ denotes the mean free time of the particle,
and $V$ is the flow speed of the scattering centers
measured in the shock rest frame.
If we adopt the Bohm diffusion model,
the mean free time of particles is given by $\tau_0 = \tau_B$,
where $\tau_B$ is the gyration time of the particle.
For this model, the residence times are given by
$t_\mathrm{res,u} \sim 2.7 \tau_B$ for the upstream region
and $t_\mathrm{res,d} \sim 13 \tau_B$ for the downstream region.
These values are indicated by the dashed and dotted lines in Fig.~\ref{fig:crossing_p}(b).
Most protons return to the shock within 
much shorter times than those obtained from the diffusion model,
while a small fraction of protons return to the shock 
with times that are comparable to or even longer than the diffusive time for low energies ($\Ekin <500$).
These results confirm that the dominant acceleration process is not diffusive.

\begin{figure}
\plotone{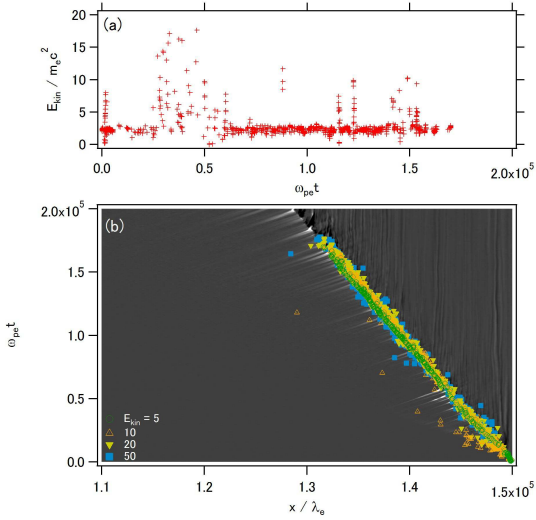}
\caption{
Injection properties for the same protons as in Fig.~\ref{fig:crossing_p}.
(a) Energies of the protons at the first crossing of the shock as a function of the first crossing time.
(b) Time and position when the protons attain the energies of $\Ekin/\mel c^2 = 5, 10, 20$, and $50$ for the first time. 
\colver
}
\label{fig:injection_protons}
\end{figure}

Figure \ref{fig:injection_protons} shows the injection properties for the same sample of protons
as in Fig.~\ref{fig:crossing_p}. The energies at their first shock crossing are shown in Fig.~\ref{fig:injection_protons}(a) as a function of the first crossing time.
Most of the protons are placed into the first shock cycle
with the upstream bulk kinetic energy ($\Ekin \sim 2.3 \mel c^2$) without pre-acceleration.
These protons will result in a constant injection of seed particles for the acceleration mechanism.
On the other hand, it is also seen within some time intervals
($3\times10^4 < \omegape t < 6\times 10^4$, $\omegape t \sim1.2\times 10^5$,
$\omegape t \sim 1.5 \times 10^5$) that the protons are pre-accelerated before the first shock crossing and gain energies up to approximately $20 \mel c^2$.
This would reflect the intermittent activity of the waves in the upstream region, which is also visible in Fig.~\ref{fig:evol_logUB}.
In Fig.~\ref{fig:injection_protons}(b), the times and the positions of the protons at which they attain energies of $\Ekin/\mel c^2 = 5,10,20$, and $50$ for the first time are shown.
We again confirm that the accelerated protons are almost constantly injected
into the acceleration process at the shock front without pre-acceleration.

\begin{figure}
\plotone{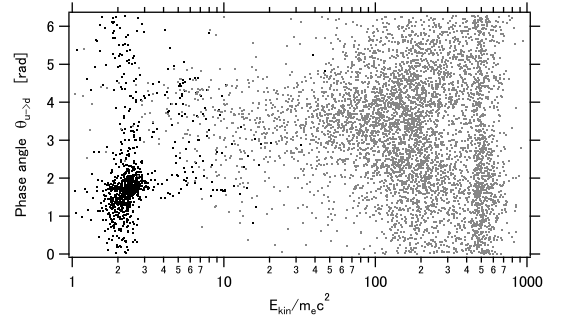}
\caption{
The angles between the transverse component of the wave magnetic field
and that of the 4-velocity of the accelerated protons
at the shock crossings form the upstream side to the downstream side.
The sample of the protons is the same as that in Fig.~\ref{fig:crossing_p}.
Those for the first shock crossing are indicated by the black dots,
and the others are indicted by the gray dots as a function of their kinetic energy at the crossing.
}
\label{fig:phase_angle_p}
\end{figure}

Figure \ref{fig:phase_angle_p} shows
the phase angles of the same sample of protons as in Fig.~\ref{fig:crossing_p}
when they cross the shock front
from the upstream side to the downstream side
as a function of the energy at the crossing.
Here, the phase angle $\theta$ is defined
as the angle between the transverse component of the wave magnetic field,
$\bvec{B}_t = (0,B_y,B_z)$, at the particle position
and that of the 4-velocity of the protons,
$\bvec{u}_t = (0, u_y, u_z)$, so that
\begin{equation}
	\bvec{\hat{B}}_t \cdot \bvec{\hat{u}}_t = \cos\theta
	\qquad \mbox{and} \qquad
	\bvec{n} \cdot \bvec{\hat{u}}_t = \sin\theta,
\end{equation}
where
$\bvec{\hat{B}}_t = \bvec{B}_t/|\bvec{B}_t|$,
$\bvec{\hat{u}}_t = \bvec{u}_t/|\bvec{u}_t|$,
and
$\bvec{n} = \bvec{\hat{x}} \times \bvec{\hat{B}}_t$.
This figure shows that, for the first shock crossings, i.e., the injection,
there is a concentration of the phase angles around $\theta \sim 1.5$.
This would show
the phase trapping of the injected protons by the upstream large-amplitude waves
\citep{Sugiyama99, Sugiyama01},
suggesting that the phase trapping can play an important role in the injection process for protons.
On the other hand, for later crossings or larger energies
the phase angles are distributed almost uniformly,
indicating that at that stage, these protons are no longer trapped by the waves
and the acceleration process becomes almost independent of the phase angle.

\subsubsection{Acceleration of Electrons}
\label{subsubsec:electron_acceleration}
\begin{figure}
\plotone{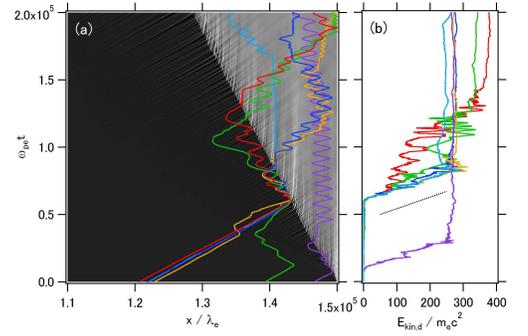}
\caption{
Acceleration histories of the highest-energy electrons
as in Fig.~\ref{fig:accel_history_p}.
As a guide,
the linear acceleration rate $d\Ekin/dt \sim 1.1 \times 10^{-2} \ (\mel c^2 \omegape)$
is shown by the dotted line in (b).
\colver
}
\label{fig:accel_history_e}
\end{figure}

\begin{figure}
\plotone{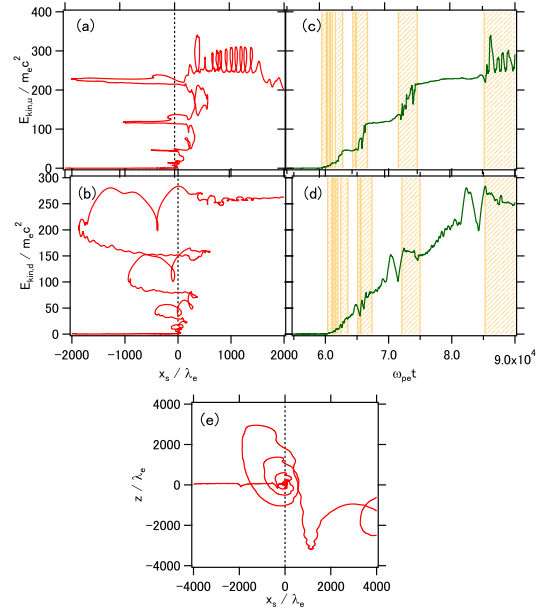}
\caption{
Representative acceleration history of electrons,
as in Fig.~\ref{fig:particle_history_p}.
\colver
}
\label{fig:particle_history_e}
\end{figure}

The acceleration process of the highest-energy electrons is shown
in Figs.~\ref{fig:accel_history_e} and \ref{fig:particle_history_e}
as in Figs.~\ref{fig:accel_history_p} and \ref{fig:particle_history_p}.
These electrons are accelerated
by essentially the same process as that of the protons.
However, there are some differences from the proton case.
The acceleration process works somewhat intermittently for the electrons.
Indeed, five of the six electrons shown in Fig.~\ref{fig:accel_history_e}
are injected into the acceleration process around $\omegape t \sim 6\times 10^4$.
The linear acceleration rate is given by $d\Ekin/dt \sim 1.1 \times 10^{-2} \ (\mel c^2 \omegape)$,
which is slightly larger than the case of the protons shown in Fig.~\ref{fig:accel_history_p}.

\begin{figure}
\plotone{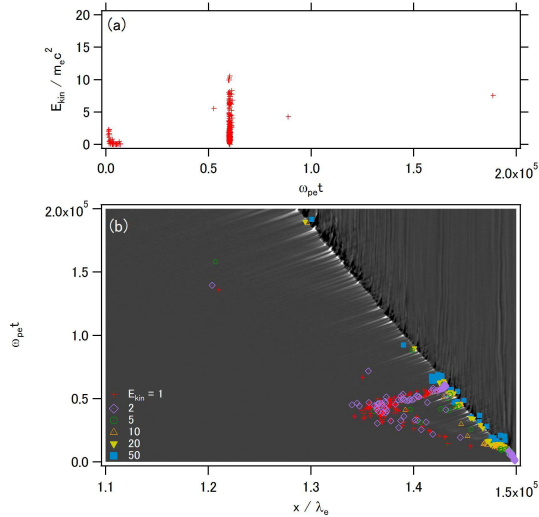}
\caption{
Injection properties for the 321 electrons that are accelerated to energies higher than $60 \mel c^2$
at the end of the simulation $\omegape t = 2\times 10^5$, as in Fig.~\ref{fig:injection_protons}.
(b) Times and positions when the electrons attain
the energies of $\Ekin/\mel c^2 = 1, 2, 5, 10, 20$, and $50$
for the first time. 
\colver
}
\label{fig:injection_electrons}
\end{figure}

\begin{figure}
\plotone{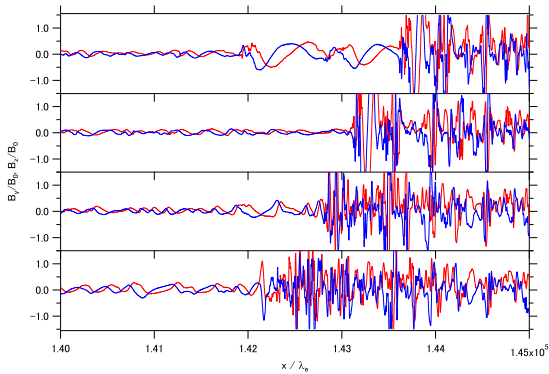}
\caption{
Evolution of the upstream waves around $\omegape t = 6\times 10^4$.
The components $B_y$ (in black (red)) and $B_z$ (in gray (blue)) are shown from top to bottom for $\omegape t = 5.5\times10^4, \ 6\times 10^4,\ 6.5\times10^4$,
and $7\times 10^4$.
\colver
}
\label{fig:upstream_wave_B_evol}
\end{figure}

Figure \ref{fig:injection_electrons} shows
the injection properties for the 321 electrons 
that are finally accelerated to energies higher than $60 \mel c^2$,
as in Fig.~\ref{fig:injection_protons}.
In contrast to the case in which
the protons are injected into the acceleration process at an approximately constant rate,
almost all of the electrons are injected
around $\omegape t \sim 6\times 10^4$, with some exceptions.
The energies at the first shock crossing are much higher than the upstream
bulk kinetic energy of the electrons (approximately $0.076 \mel c^2$),
indicating a pre-acceleration in the upstream region.
The time of the efficient injection coincides with
the end of the time interval observed in the proton injection
in which the pre-acceleration of protons is efficient
(see Fig.~\ref{fig:injection_protons}(a)).
During this time interval,
the amplitudes of the upstream waves become very low 
(see Fig.~\ref{fig:upstream_wave_B_evol}).
On the other hand,
just before that time, the wave amplitude grows to very high level.
The injection efficiency of both protons and electrons
would be related with these wave activities.

\begin{figure}
\plotone{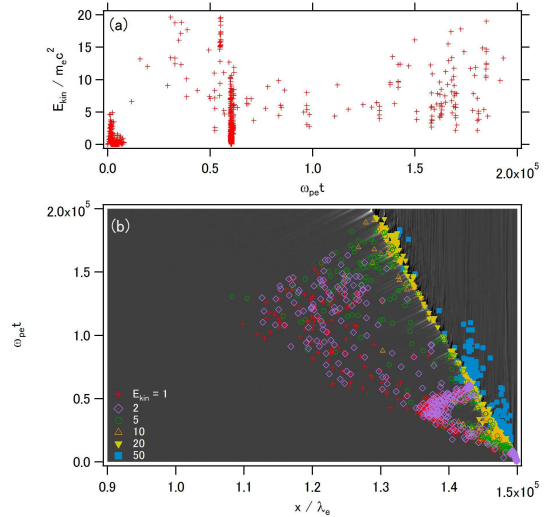}
\caption{
Same as Fig.~\ref{fig:injection_electrons}, but for
the 1,000 electrons accelerated to energies within the range of between $50 \mel c^2$ and $60 \mel c^2$
at $\omegape t = 2\times 10^5$.
\colver
}
\label{fig:injection_electrons_2}
\end{figure}

The injection properties for the 1,000 sample electrons that achieve slightly lower energies $50 < \Ekin/\mel c^2 < 60$ at the end of the simulation is shown in Fig.~\ref{fig:injection_electrons_2}, as in Fig.~\ref{fig:injection_electrons}. In this case, the electrons are injected into the acceleration process even at other times. The electrons are pre-accelerated in a part of the upstream region and the width of this region is extended with time, as shown in Fig.~\ref{fig:injection_electrons_2}(b). This region coincides with the foreshock region, where large-amplitude waves exist (see Fig.~\ref{fig:evol_logUB}). With this extension of the pre-acceleration region, the typical energies at the first shock crossing are also increasing, as shown in Fig.~\ref{fig:injection_electrons_2}(a) for $\omegape t > 1\times 10^5$. In particular, for $\omegape t > 1.5 \times 10^5$, approximately half of the injected electrons have gained energy in the upstream region to higher than $10 \mel c^2$ before the electrons reach the shock front.
Although these pre-accelerated electrons have not contributed to the highest-energy electrons shown in Fig.~\ref{fig:injection_electrons} by the end of the simulation,
they may contribute at a later time.

\begin{figure}
\epsscale{0.8}
\plotone{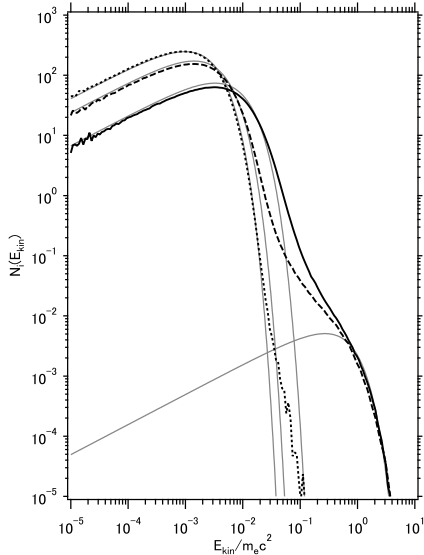}
\caption{
Electron energy spectra in three upstream regions at $\omegape t = 1\times 10^5$,
where the shock is located at $x \sim 1.39\times10^5 \lambdae$
and the energies are measured in the upstream rest frame.
Electron energy spectra obtained for the regions of $9\times10^4 < x/\lambdae <9.5\times10^4$ (dotted curve), $1.1\times10^5 < x/\lambdae <1.15\times10^5$ (dashed curve),
and $1.3\times10^5 < x/\lambdae <1.35\times10^5$ (solid curve)
are shown in black. The curves in gray show the fitted relativistic Maxwellians
with temperatures of $T/\mel c^2 = 0.0020$ (which is the far upstream temperature),
$0.0028$, $0.0065$, and $0.35$.
}
\label{fig:hist_u_100000_e}
\end{figure}

Figure \ref{fig:hist_u_100000_e} shows 
the energy spectra of the electrons
calculated for three regions
upstream of the shock
at $\omegape t = 1\times 10^5$.
The bulk upstream electrons are heated 
by advection as they approach the shock.
Their energy spectra in each region are well fitted by Maxwellian distributions.
Moreover, a fraction of electrons are accelerated to high energies
in the foreshock region, where the amplitude of the upstream
waves becomes large, as shown in Fig.~\ref{fig:injection_electrons_2}.
This component can also be fitted by a Maxwellian
although its temperature (approximately $0.35 \mel c^2$)
is much higher than that of the bulk electrons.
As already mentioned,
these pre-accelerated electrons would result in
the high-temperature Maxwellian component in the downstream region.
These pre-acceleration processes may later contribute to the steady electron injection
for the Fermi-like acceleration process.

\section{Discussion}
In the present simulation of a quasi-parallel shock,
we observed ion (proton) acceleration
and the associated wave generation around the shock.
Such processes are also observed in quasi-parallel shocks in the heliosphere
by \textit{in situ} observations with spacecraft.
In particular,
the ion acceleration is observed in,
for example,
the Earth's bow shock \citep{Burgess12}
and the shocks associated with the solar energetic particle events \citep{Lee12}.
For the latter,
the ion energy spectrum shows a power-law shape
and considered to be accelerated by the diffusive first-order Fermi acceleration mechanism.
On the other hand,
in the present simulation,
as shown in Section \ref{subsubsec:proton_acceleration},
the dominant acceleration mechanism are not diffusive
but a process that has a shorter acceleration timescale,
which is similar to the ``scatter-free'' acceleration mechanism \citep{Sugiyama01,Sugiyama11}.
Regardless of the difference in the dominant acceleration mechanism,
the resulting energy spectrum still becomes a power-law shape.
The reason would be that it still satisfies the fundamental requirements
for the Fermi-type acceleration mechanism,
i.e., the repeated shock crossings, the elastic scattering in the scattering center rest frame,
and an approximately constant escape probability from the acceleration cycle,
although the spatial motion of the particles is not diffusive.
The large-amplitude upstream waves observed in the simulation
have a similar structure to the ULF waves observed
in the quasi-parallel Earth's bow shock by \textit{in situ} observation \citep{Hoppe81};
in the present simulation,
the reflected/accelerated protons would play a similar role
to the energetic diffuse ions in Earth's bow shock
for generating these waves.
The somewhat regular perpendicular magnetic structure due to the generated waves
in the upstream region observed in the simulation
may be caused by the 1D dimensionality of the simulation.
In fact,
recent multi-dimensional hybrid simulations of
high-Mach-number quasi-parallel shocks 
show more turbulent magnetic structures around the shock transition region \citep{Caprioli14}. 
To see whether the 1D dimensionality affects the results significantly,
some multi-dimensional simulations are desired.

\citet{Amano10} proposed an electron injection model with
a critical Alfv\'{e}n Mach number
above which the electron injection occurs.
From observations in the heliosphere,
it seems that this criterion is satisfied in several shocks
in which electron acceleration is observed \citep[e.g.,][]{Oka06,Masters13}.
It would be worthwhile to see
whether this electron injection mechanism operated in the present simulation.
The electron acceleration occurred only around
$\omegape t \sim 6 \times 10^4$ in the simulation.
In this time interval, the amplitude of the upstream waves
is temporally reduced as shown in Fig.~\ref{fig:upstream_wave_B_evol}
and the local inclination angle of the magnetic field at just upstream of the shock
is given by $\Theta \sim 30^\circ - 60^\circ$.
The critical Mach number is thus given by $M_A^\mathrm{inj} = 0.8 - 1.5$.
On the other hand, the local Alfv\'{e}n Mach number is only slightly
reduced from the global value of $M_A=28$.
Hence, the condition for injection $M_A \gg M_A^\mathrm{inj}$ is satisfied
and, from the criterion argument, the electron injection is expected to occur.
In their injection model, it is assumed that a fraction of the incoming electrons are reflected
at the shock front to form the electron beam and then generate the whistler waves
in the upstream region,
and finally these waves scatter the electrons leading to the diffusive acceleration.
Thus, we seek such electron beams reflected at the shock front
in the phase-space plots at $\omegape t=4\times 10^4$, $5\times 10^4$, and $6 \times 10^4$.
However, such beams couldn't be found. In addition, the trajectories of the accelerated
electrons show no clear indication of the interaction with the whistler waves.
Therefore, it seems that the injection mechanism of the electrons observed
in the simulation is different from those considered in \citet{Amano10}.
On the other hand,
in the other times,
their injection mechanism does not operate, too,
although the criterion is still satisfied,
where
the local inclination angle of the magnetic field is typically given by $\Theta \sim 80^\circ$
and the critical Mach number becomes $M_A^\mathrm{inj}<1$.
Regarding this point,
it should be noted that
in the present simulation the superluminal condition can be realized
because the shock velocity is relatively large ($\sim 0.46c$)
and also the local inclination angle of the magnetic field becomes quasi-perpendicular
after the upstream waves grow substantially;
the condition for the shock to be subluminal is $\Theta < 63^\circ$ for $V_s = 0.46c$
and this is difficult to be satisfied after the upstream waves grow to large amplitude.
In such cases, the de Hoffmann-Teller frame does not exist and so
the incoming electrons cannot be reflected at the shock front.
Thus, to see whether the electron reflection at the shock front occurs and the injection process proposed
by \citet{Amano10} operates, simulations with smaller shock velocities would be needed.
This should be investigated in the future studies.

In Section~\ref{subsection:shock_structure}, the local shock structure becomes substantially quasi-perpendicular after the amplitude of the upstream waves grow to be sufficiently large. This would be a common feature of the quasi-parallel shocks in which the particle acceleration is efficient and the accelerated particles excite large-amplitude waves in the upstream region. In such local quasi-perpendicular conditions, as observed in several simulations of the perpendicular shocks \citep{Amano09, Kato10, Sironi11, Riquelme11, Matsumoto13}, the electron heating and/or acceleration in the foot region of the shock structure can also occur (depending on the shock parameters and the dimensionality of the simulation) in addition to the electron acceleration/heating in the upstream wave region found in Section~\ref{subsubsec:electron_acceleration}. Regarding this shock structure, while we observe that the shock is simply formed in the environment of the local quasi-perpendicular magnetic field, the shock formation due to the nonlinear steepening of the upstream waves themselves was also reported for relativistic parallel shocks \citep{Sironi11}. This would indicate that the shock structure can depend on the shock speed. Which shock structure is realized would be determined by whether the typical gyro-radius of  the protons reflected at the shock front is larger than the typical wavelength of the upstream waves.

\begin{figure}
\epsscale{0.8}
\plotone{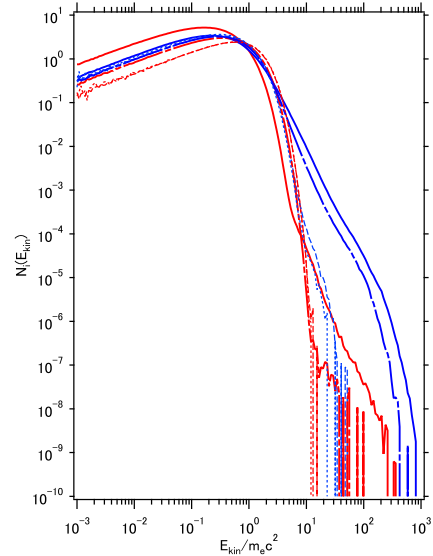}
\caption{
Energy spectra in the downstream region of the shock at $\omegape t = 2\times 10^5$ for protons (in black (blue)) and electrons (in gray (red)), as in Fig.~\ref{fig:hist_d}, for four simulations that are identical except for the number of super-particles within the electron skin depth, $\Ne$: $\Ne=50$ (dotted curves), $\Ne=133$ (dashed curves), $\Ne=667$ (dot-dashed curves), and $\Ne=2,667$ (solid curves), for which the results are shown in Section~\ref{sec:results}.
\colver
}
\label{fig:hist_N_dep_4000000}
\end{figure}

Note that, for PIC simulations, in particular those dealing with the particle acceleration process, as in the present work, the number of super-particles used in the simulations can be important because high-energy particles in the simulations generally suffer from the energy loss process due to the stopping power of the plasma \citep{Kato13}. The energy loss rate is inversely proportional to the number of super-particles in the electron skin depth volume, $\Ne$. Therefore, if $\Ne$ is too small, the energy loss process becomes significant making the acceleration process inefficient. For one-dimensional simulations, the energy loss rate for relativistic particles is given by
\begin{equation}
	\frac{d \Ekin}{d t} \sim -\frac{1}{2 \Ne} (\mel c^2 \omegape).
\end{equation}
From Figs~\ref{fig:accel_history_p} and \ref{fig:accel_history_e}, if we take the representative value of the acceleration rate for the present case to be $d\Ekin/dt \sim 1\times10^{-2} (\mel c^2\omegape)$, the value of $\Ne$ for which the energy loss rate is equal to the acceleration rate, is given by $\Ne \sim 50$. Figure \ref{fig:hist_N_dep_4000000} shows the energy spectra of protons and electrons in the shock downstream region at $\omegape t = 2\times 10^5$, as in Fig.~\ref{fig:hist_d}(a), for four simulations that are identical except for the value of $\Ne$. The acceleration efficiency is indeed dependent on $\Ne$ for both protons and electrons. In particular, in the case of $\Ne=50$ (in which the number of super-particles per cell is $\NPPC = 3$ for $\Delta x = 0.06\lambdae$) and $\Ne=133$ ($\NPPC = 8$), the acceleration process becomes significantly inefficient and is almost completely nonfunctional. Even in the case of $\Ne=667$ ($\NPPC=40$), the acceleration efficiency is still affected by the energy loss. Thus, for these simulations, the number of super-particles used should be chosen to be sufficiently large, so that the energy loss is negligible.

In the present paper, we observed that electron injection occurs only within a particular time interval ($\omegape t \sim 6\times 10^4$) and does not settle at a constant rate, at least during the calculation time of the simulation. For longer timescales, the electron injection would be able to occur constantly or repeatedly. Instead, the electron injection observed in the simulation may be a consequence of the initial condition and may not occur again. In order to resolve this issue, further long-term simulations are required. In addition, in two or three dimensions, some electromagnetic instabilities, such as the ion-Weibel instability, can play important roles, even for non-relativistic shocks \citep{Kato08, Niemiec12}, and these effects can influence the particle acceleration process and the shock structure. Thus, multi-dimensional and long-term PIC simulations are also desired.

\section{Conclusion}
We have carried out a long-term, large-scale PIC simulation of a quasi-parallel high-Mach-number shock in an electron-proton plasma. We showed that both protons and electrons are accelerated in the shock, and these accelerated particles generate large-amplitude Alfv\'{e}nic waves in the upstream region of the shock. The local structure of the collisionless shock becomes substantially similar to that of the quasi-perpendicular shocks because the transverse components of the incident upstream waves dominates the parallel background field. A fraction of the protons are accelerated in the shock with a power-law-like energy distribution as has been demonstrated through several hybrid simulations. The injection process for the protons operates almost constantly. In the process, phase trapping of the protons by upstream waves can play an important role, while the later acceleration process is nearly independent of the phase angle. The dominant acceleration process is a Fermi-like process that occurs through repeated shock crossings of the protons but is not diffusive. Most of the accelerated protons complete one-cycle of the acceleration process within a time on the order of the gyration time, which is much shorter than the diffusion time. We also found that electrons are accelerated in the shock by the same mechanism, and the energy spectrum of the accelerated electrons has a power-law like distribution. However, the injection is not constant, and electrons are actually injected during only one time interval in the simulation. This behavior would be related to the intermittent activity of the upstream waves. Upstream of the shock, a fraction of the electrons is accelerated before reaching the shock, which would result in a two-temperature electron distribution in the downstream region. At a later time, the pre-acceleration process may contribute to steady electron injection for the Fermi-like acceleration process.

\acknowledgements
We would like to thank the anonymous referee for the valuable comments.
This work was supported by JSPS KAKENHI Grant Number 22740164, 24540277.
Numerical computations were carried out on Cray XC30
at Center for Computational Astrophysics,
National Astronomical Observatory of Japan.



\end{document}